\documentstyle[12pt]{article}

\def\title{\bgroup\obeylines\everypar={\hskip\parfillskip}\large
	   \bf\vrule height1cm width 0pt\relax}
\def\endtitle{\vskip1sp\egroup}
\def\author#1{\hbox to\textwidth{\hss\vrule height.9cm width0pt\relax
 #1\hss}}
\def\contauthor#1{\hbox to\textwidth{\hss\vrule width0pt\relax #1\hss}}
\def\moreauthors#1{\hbox to\textwidth{\hss\vrule height.8cm
		   width0pt\relax #1\hss}}
\def\instit{\bgroup\small\it\obeylines\everypar{\hskip\parfillskip}}
\def\endinstit{\vskip1sp\egroup}
\def\infap{<\mkern-19mu{\lower1.3ex\hbox{$\sim$}\;}}
\def\supap{>\mkern-19mu{\lower1.3ex\hbox{$\sim$}\;}}

\newcommand{\be}{\begin{equation}}
\newcommand{\ee}{\end{equation}}
\newcommand{\bea}{\begin{eqnarray}}
\newcommand{\eea}{\end{eqnarray}}

\begin{document}
\bibliographystyle{my}

\vglue -4 true cm
\vskip -4 true cm
\def\today{ June 1995}

\begin{center}
{\hfill }{ FTUAM/95-23}
\end{center}

\vskip 1.5 true cm

\begin{title}
Gauge invariant structures and Confinement \footnote{Accepted for publication  in Physics Letters B.}
\end{title}

\medskip
\begin{center}

A. Gonz\'alez-Arroyo, P. Mart\'{\i}nez \footnote{Present adress: Dept. of Physics, The University of Edinburgh, Edinburgh EH9 3JZ, Scotland.} and A. Montero 

{\it Departamento de F\'{\i}sica Te\'orica C-XI, Universidad Aut\'onoma
 de Madrid,
\\ 28049 Madrid, Spain.}

\end{center}

\begin{center}
\today
\end{center}
\medskip
\begin{abstract}
By looking at cooled configurations on the lattice, we study the presence of peaks in the action density, 
or its electric 
and magnetic components, in the SU(2) gauge vacuum. The peaks are seen to be 
of instanton-like nature and their number variation takes care of the drop
in the string tension observed when cooling. Possible explanations of this finding  are 
analysed.
\end{abstract} 	

\newpage

\section{Introduction}
Our purpose  in this  paper is to present results aiming at the understanding of the
Confinement mechanism in the gauge theory vacuum.  We will  study SU(2) 
 Yang-Mills theory for simplicity reasons, but it is generally believed
that there is a common origin for the Confinement problem for all gauge groups and 
in the presence of quarks. Quantitative differences are nonetheless expected, 
about which we have nothing to say at present. In seeking for a simple understanding 
of the Confinement problem, many workers in the field have pursued the relationship 
of the non-abelian theory with an abelian theory with charges and monopoles, which 
is known as the abelian projection \cite{thooft6}. Confinement is then seen as 
dual superconductivity \cite{thooft1,mandel1} arising from  the condensation of magnetic charge. 
Despite the appealing aspects of this approach, there are still some aspects which should 
be clarified. For example, the class of acceptable gauge fixing projections has to be
understood, and the relationship between different gauges clarified (see for example
\cite{chernodub}). Here, we will pursue a different avenue, not necessarily incompatible with  the 
previous one. We will  look for gauge invariant structures which  might  reveal 
to be associated with Confinement. Their existence is a common feature of all models of Confinement 
based on classical Yang-Mills configurations. In our prefered scenario, the gauge-invariant
 structures would,
after the abelian projection, give rise to monopole currents.  

Now we should proceed to analyse the possible methodology. Monte Carlo generated lattice 
 configurations are 
seen to reproduce the statistical properties of the Yang-Mills vacuum and give rise to a
non-zero string tension. However, it is very hard to see any structure directly in these
configurations, since the short wavelength noise is much bigger than any signal. 
A traditional way out of this problem is to make use of the technique known as 
cooling \cite{cool1,cool2}. This local minimization mechanism is intended to relax the noise at a 
much faster rate than the long wave-length signal.  This strategy was already clear  in
 Polikarpov and Veselov work \cite{polikarp}.  Since then, other authors have pursued 
similar roads \cite{negele1,gupta,michael2,michael3,trottier}.

 A fundamental question is to understand the effect of cooling on the
resulting  configurations. One should make sure that what one sees in the end is not an artifact 
of the cooling procedure, and if it is not, that there is no distortion introduced by this
procedure. Fortunately, cooling is not a unique technique, and there
are several versions of it,
 which allow to test the universallity of the results. In particular,
we have found convenient to consider a 1-parameter family of cooling procedures, which we will
refer to as overimproved cooling \cite{uscool}. 
 This is nothing else but the naive cooling method, but associated with a
one parameter ( $\epsilon$) family of lattice actions formed by $1 \times 1$ and $2 \times 2$
plaquettes. The advantage, is that in going from the Wilson action ( $\epsilon   = 1$)
to $\epsilon   = - 1$, one has reversed the sign of the $O(a^2)$ corrections to the
continuum action. 
 The point is quite relevant, since order $a^2$ terms break the degeneracy
of the zero-modes and induce distortion of the resulting cooled configuration. In
particular, the well-known instability of lattice instantons under cooling can be seen to
be a result of  the wrong  sign (positive) of  $\epsilon$. Choosing negative values of 
$\epsilon$ is hence recommended. Nonetheless, the larger the absolute value of
$\epsilon$, the faster the motion along zero-mode directions of the original
configuration, with the corresponding modification of the original distribution  
 ( instanton size distribution for example).  For our lattice sizes and values of 
$a$,  $\epsilon= -0.3$ is a good compromise. A fairly complete analysis of the effect of cooling 
on the lattice configurations for different values of $\epsilon$ has been presented
elsewhere \cite{uslat94,uscoolnew}. 

A mechanism which results from cooling, but is fairly 
independent of the value of $\epsilon$,  is  the annihilation of instanton-antiinstanton pairs. 
Although preserving the value of the topological charge,
this mechanism induces a reduction of the number of instantons with cooling.   
 The typical number of cooling steps necessary 
to produce an annihilation, depends very much on distance and width of the 2 intervening instantons,
but it is not untypical to have destructions within a few tens of cooling  steps. 
We do  not know at present if it is possible to devise a different technique 
 which is free of this problem. Hence, we should understand at least how this mechanism 
affects our results.

\section{Analysing the gauge invariant content of the vacuum}
Our analysis is based on  SU(2) Yang-Mills configurations
obtained by Monte Carlo simulations with Wilson action  and $\beta = 2.325$
in lattices of size $8^3 \times 64$. We imposed twisted boundary 
conditions in all spacelike planes, corresponding to twist vector $\vec{m} = (1,1,1)$. In total
 we have used 56 configurations separated by several thousand
heat bath sweeps on the average. Our estimates show that the configurations are fairly statistically
 independent from each other. The choice of the point and
parameters of the simulation was done based on our previous analysis
\cite{uslat93,uscool2}. In these previous works, we studied the electric flux ground state energies 
by measuring the temporal  correlation between Polyakov loops of different winding numbers in the (short)
spatial directions. Our results showed  a fairly good scaling behaviour for these
quantities even for  $N_s=4$: The energies depended on  the lattice spatial length  $N_s$
and the plaquette coupling constant $\beta$, only through  the physical length $l_s = N_s a(\beta)$.
 Furthermore, we had  a fairly extensive coverage of the dependence 
of the energies with $l_s$, showing that, at values of $l_s$ slightly exceeding $1 fm$, the results 
obtained on these intermediate  lattices were consistent with both the Confinement prediction and the results
of much larger lattices. More specifically, what we had is that the electric flux ground state energies
$E_{\vec{e}}$ behaved as 
\be
E_{\vec{e}}= | \vec{e} | \sigma \l_s
\ee 
where $\sigma$ is the string tension. At $N_s=8$ and $\beta=2.325$, the values of $\sigma$ obtained for the three different fluxes
$\vec{e}=(1,0,0)$, $\vec{e}=(1,1,0)$ and $\vec{e}=(1,1,1)$ are $4.75(17) fm^{-2} $, $4.46(44) fm^{-2} $
 and  $5.01(30) fm^{-2}$ respectively.
We see that the value is consistent for the three  determinations and also 
consistent with the large volume result, which is taken for normalization, equal to $5 fm^{-2}$.
The $l_s$ dependence in this range of volumes is also seen to be consistent with Formula (1).
From these results we conclude, that at this value of $\beta$ and this size, we are in fact 
measuring  the string tension with fairly small finite 
size corrections (  10 \% ). 

We have performed  up to 50 cooling steps on  our configurations with our over-improved ansatz and $\epsilon= -0.3$.
One can measure the electric flux energies   $E_{\vec{e}}$  on the cooled configurations,
and from them extract a value of  $\sigma$ by means  of Eq.  (1). Again, the three values obtained from the
different fluxes are consistent, but these values are now smaller than the ones obtained before 
cooling and with the full statistics. Furthermore, the  value of  $\sigma$ decreases with the number of cooling steps.
Our results were presented in Ref. \cite{uslat94}, where we pointed out that at 50 cooling steps, the value 
of the string tension had only dropped around 40-50 \% of its uncooled value. Given the degree of self-duallity 
observed in these cooled configurations, we argued that a big fraction, if not all, of the physical string tension
is due to classical (smooth and almost self-dual) configurations. We knew not, however, what could be the origin of the
 observed decrease with cooling steps. 

 Our results, and in particular the afore-mentioned decrease, is apparently in
contradiction with some  statements made previously by other authors 
 on similar issues. For example, in Ref. \cite{polikarp} the authors claimed a much smaller fraction of the 
string tension to be attributed to instantonic configurations. Nonetheless, there are differences between their work 
and ours in several respects: These authors  cool for a longer period and with the naive cooling method, which given the tendency to
 decrease, might well explain the differences. Another work going in the opposite direction is that of Ref. \cite{digiacomo}.
The authors claim that the string tension becomes stable under cooling, but with very rapidly decreasing errors. 
The apparent disagreement with our results, could be understood if we take into account that again the authors use a
different cooling method but this time they cool for a shorter interval than us. Actually, in their work they acknowledge an eventual
drop of the string tension beyond some number of cooling steps. To understand  what could be the origin of the difference,
we tried using their cooling procedure instead of ours. What we observed, is that the cooled trajectory seems to be the same 
with both cooling methods, but that every ordinary cooling step    corresponded to several  of their steps. Indeed, our results 
show a plateau for the first 3 to 5 steps, resulting in a longer one for the method
of Ref. \cite{digiacomo}.
Presumably, hence, there is no inconsistency between our results and those of other authors. Finally, we should mention the remark
made by Teper \cite{teperarg} that, since cooling is a local operation, all masses including the string tension have to remain invariant under cooling.
This result, although correct, still permits the appearance of a mass parameter at intermediate length scales, which is the one we are referring to.
We claim that this new string tension exists and has a physical meaning. Indeed, since cooling rapidly brings lattice configurations to smooth ones,
its following effect can be understood in purely classical terms. The afore-mentioned instanton-antiinstanton annihilation mechanism for example,
can only take place, in a finite number of cooling steps, among pairs that are separated less than a certain distance. Only beyond this scale can one recover 
the asymptotic invariant string tension. In summary, the results of our  paper, when taken in this context, are both meaningful and 
consistent with Teper's argument.  

In this work we have looked at the structure and content of cooled configurations. We have concentrated in local maxima 
of the action density and of its electric and magnetic components. We also triggered on similar maxima for the absolute value 
of the topological 
charge density. Our definition of the lattice densities is as follows. We first construct  the colour electric and magnetic field
 at each lattice point, by averaging over all plaquettes emerging
from the point in question in the appropriate plane, in a clover-like fashion.  We also 
used a naively improved version, which uses both $1 \times 1$ and  $2 \times 2$ plaquettes (clover=2).
Given the fields, the densities are obtained by scalar 
products in colour space and ordinary space according  to the continuum formulas.
Next, our algorithmic definition of a maximum, is given by the points where the value of the quantity in question is larger than in all its first neighbours. 
We call such a maximum:  a peak. For each peak, 
its  position and the value of the quantity under consideration at this point and at its nearest neighbours is stored. 
 It is possible to  calculate its position with better than a lattice spacing precision, by fitting a parabolid with the afore-mentioned data 
and extracting its maximum. This gives excellent results on latticized classical configurations usually.   

Our first result concerns the global statistics on the number of observed peaks. First, we mention that there is no difference
observed between the number of electric and magnetic  peaks (There is a tendency to have a higher number of 
peaks in the topological charge density, due to noise caused by it being a signed quantity). Not only the number 
of peaks agrees, but also the location and the size is the same in more than 80 percent of the cases. We are hence dealing with 
self-dual structures associated with the peaks. We expect that instantons are present and cause some of the observed peaks. 
In the continuum, the action density  associated with an instanton is given by:
\begin{equation}
S(x)= \frac{48\rho^4}{{(x^2 + \rho^2)}^4} \ .
\end{equation}
Since $\rho$ is the only scale  present, both the height and the width are functions of it. This suggests considering 
the following two functions of the peak structure:
$$
\rho_1= {\left( \frac{48}{S(0)}\right) }^{\frac{1}{4}}
$$
\begin{equation}
\rho_2= \sqrt{\frac{-32S(0)}{\triangle S(0)}} \ .
\end{equation}
Notice that, while the first quantity is only sensitive to the height of the peak and not to the width, the second is
invariant under a scale transformation. They are, hence, quite independent from each other 
in general. However, for an instanton both quantities are equal to the size parameter $\rho$.  For the  set of our sample
peaks we have computed $\rho_1$ and  $\rho_2$ by using the information of the peak and its nearest neighbours in an obvious way.
In Fig. 1 a contour plot is given which shows the combined distribution of both variables for the 50
 cooling steps  data. No 
essential difference is obtained by looking at 25 cools, electric or magnetic and clover 1 or 2.
The message of Fig. 1 is that our peaks are locally similar to instanton peaks. This behaviour is not a general consequence 
of self-duallity and the existence of a peak. The range of observed values of $\rho_1$ is determined by our ultraviolet
and infrared cut-offs. We can produce instanton-like configurations with values of $\rho$ close to the extremes observed
in the contour plot. We cannot claim, nevertheless, that our peaks are BPST instantons, since there are other configurations 
showing a similar structure at the center of the peak but differing away from it. For example, that 
is the case for the $Q=\frac{1}{2}$ instanton
observed with twisted boundary conditions \cite{us1,jouphys}.

 The observed number of peaks decreases  with the  number of applied cooling steps. 
Is there any correlation between this decrease and the one of the string tension? On very general grounds 
one can relate both quantities to a scale. The number of peaks per unit volume gives  a certain density which has dimensions
of length to the power -4. The string tension has dimension of length to the power -2. Hence, we may form the scale invariant 
combination
\begin{equation}
 K = \frac{\sigma}{ \sqrt{N_{peaks} / Volume}}
\end{equation}
 Being dimensionless, it can be computed both in lattice or in physical units.
In Fig. 2 we show the value of K as a function of the number of cooling steps. We see that after 5 coolings, the value of K 
shows a striking plateau. Although errors are shown, one must take into account that data at different number of cooling steps 
are strongly correlated. Hence, the plateau is much better that what errors might allow.  In the same plot the data of the string 
tension (divided by 2 in $fm^{-2}$ units) is shown, to let the reader appreciate how the computation of the quotient does correct the decreasing behaviour. 
The quality of the data leaves no doubt that the drop in the number of peaks and in the string tension have the same origin.

The extraction of the string tension from our lattice data is done by measuring correlations of straight-line Polyakov loops $C(t)$ at time separation $t$. 
 As usual, the corresponding mass is  obtained by considering first the effective mass defined as 
\begin{equation}
M(t) = - ln \left( \frac{C(t+1)}{C(t)} \right) \ .
\end{equation}
The mass parameter ( which divided by $l_s$ gives the string tension) should be seen as a plateau in the value of M(t) for suficiently  large t. Our data values
for M(t)  do grow
 with $t$ for small values, but show a levelling up for 
distances $t$  of 5 or 6. Unfortunately,  $t$ cannot exceed values of 6 or 7 since the errors grow too much. Combining this data with the definition 
of K, we might define $K(t) = M(t)  / ( l_s \sqrt{N_{peaks} / Volume})$. The plateau behaviour is obtained with similar quality for all  values of $t$,
except for 5 coolings for which we have a better behaviour for large $t$. From the results obtained from 20 to 50 cooling steps our plateau values obtained are:
 $K(2)=0.939 \pm 0.024 \pm 0.027$,
$K(3)=1.463 \pm 0.048 \pm 0.037$, $K(4)=1.881 \pm 0.097 \pm 0.036$, $K(5)=2.184 \pm 0.205 \pm 0.027$, $K(6)=2.331 \pm 0.427 \pm 0.028$,
$K(7)=2.236 \pm 0.813 \pm 0.048$.  The first error is statistical and is basically equal for all cooling steps.  The second is the maximum difference  observed
in absolute value between the average and  the value for all cooling
steps from 20 to 50 . It is a measure of the quality of the plateau. We stress that while the number of peaks drops by a factor of 2 from 20 to 50 cooling steps,
K varies by less than 2\% in this interval.

\section{Results and Conclusions}
The results shown in the previous section demonstrate the existence of  a clear 
correlation between the value of the string tension and the occurence 
of peaks in the action density, or its electric and magnetic components.
Furthermore, we have shown some evidence that the observed peaks are 
self-dual and of instanton-like character. Can we understand how this can occur?
As mentioned previously, the quantity which shows a plateau is  a dimensionless
combination. Hence, we can explain such a behaviour if we consider a  theory 
with a single length scale which could enter both the density of "objects"
and the string tension. Furthermore, in order to explain the drop in the string 
tension, it should happen that cooling produces 
a growth of this scale. We only know of one model that predicts such a behaviour,
although there could be others as well. In this model, proposed by some of us
\cite{uslat93,uscool2}, Confinement is a property of a certain class 
of multi-instanton configurations, which are argued to dominate the path 
integral (with their cloud of perturbative fluctuations). The set of configurations is
 conjectured  to be describable as a 4 dimensional liquid of $Q=\frac{1}{2}$ instantons.
 The main parameter of this fluid is its density, which is fixed dynamically
to a value that was estimated to be in the order of $(0.7 fm)^{-4}$. It was essentially this 
value which set the scale to  the string tension, which was estimated to be 
roughly of the size known phenomenologically. In what follows we will show that our data 
are in other respects in agreement with the predictions of this model.

 Obviously one of the  characteristic ingredients of the model of Ref. \cite{uslat93,uscool2}
is that the objects which make up the gas are self-dual and have fractional 
topological charge. Unfortunately, as mentioned previously, it is not possible to distinguish
$Q=\frac{1}{2}$ instantons from $Q=1$ instantons by looking in the  neighbourhood of the maximum. 
By now  we have been unable to find a simple algorithmic procedure to distinguish
both types of configurations. We might however proceed in an statistical fashion and 
compute the mean action divided by $4 \pi^2$ per peak configuration by configuration. This quantity also turns out to be
fairly insensitive to cooling, varying from 1.072 to 1.132 as we move from 20 to 50 cooling steps.
A more  informative plot is that of Fig. 3,  where we have plotted the ratio $\frac{S}{ 4 \pi^2}$ versus the 
number of peaks $N_{peaks}$
for every configuration and for all cooling steps from 20 to 50 ( in steps of 5). The two displayed lines correspond to 
the predictions for a gas of instantons (slope 2) and of 
$Q=\frac{1}{2}$ instantons (slope 1). Clearly all points cluster around the second line. 

That the mean action per  peak is equal to $4 \pi^2$ doesn't necessarily mean that all our peaks 
are  $Q=\frac{1}{2}$ instantons. For example if some peaks carry no action and others have the typical
instanton action $8 \pi^2 $, one can still end up getting the same mean action. To convince ourselves 
that this is not what is happening, we have been monitoring individual configurations as we cool the system.
A good advantage of our elongated lattice shapes ($64 \times 8^3$),
is the nice aspect of  the  energy profile,  defined as 
\begin{equation}
{\cal E}(t)  =  \int d^3\vec{x} S(x)  \ .
\end{equation}
As observed previously \cite{us1},  this quantity behaves very approximately as if it was additive. If we have a collection
of  $Q=\frac{1}{2}$ instantons centered at different times, we might simply add the known profile of each 
object to have a prediction for the total profile. In some cases we need to add an ordinary  $Q=1$ instanton
profile with a given width $\rho$ to have a good description. Fig 4. shows the comparison of the lattice data
with the prediction for one of our  configurations at 240 cooling steps.  Essentially identical results are gotten 
 for 130 cooling steps. The discontinuous curve is obtained by adding the energy profiles of a set of $Q=\frac{1}{2}$ instantons
centered at the time positions where we have observed peaks in the action density ( no free parameters).
To describe the data better, 3 of the peaks are set to $Q=1$ instantons with values of $\rho$ which agree
with the  $\rho_1$ or $\rho_2$ parameters of the peak. If we leave the positions and widths as free 
parameters, we can obtain  a line which goes basically through all our data points, which is also shown in  Fig 4.
The new locations of the instantons are in most cases within one lattice spacing of the peak position. The 
displacement is  larger in those places where there are other nearby peaks, as expected.
This case shows that indeed, at least for this configuration and cooling value,
 the majority of our peaks are associated  with $Q=\frac{1}{2}$ instantons. The other configurations that we have 
looked at show essentially the same behaviour.

Now suppose that indeed the ratio K is the same all the way down to zero coolings. Of course it is very hard
to count the number of peaks for zero cooling steps, since most of the peaks are purely the result of 
high  frequency noise. This might explain also why at 5 cooling steps there is a drop in the measured 
value of K: together with the instantonic peaks there are also some noise peaks.
 Notice nonetheless that at 5 cooling steps one can extract values of the string tension which 
are fairly in agreement  with the uncooled ones and that the departure of K from the plateau value there is not too big.
Hence, we might be seeing indeed a property of the uncooled vacuum.  Since the string tension value is known
( $\sigma = 5 fm^{-2}$), we can use the measured value of K to predict the density. 
A fairly safe estimate for K, coming from $t=6$ is $K=2.33(43)$  , which gives for the 
density $(0.68 (7) fm)^{-4}$. Notice that the resulting value is very close to the estimated 
density for our model.

Finally, let us give  our explanation for the decrease of the number of peaks or of the string tension. 
Indeed, if those peaks were instantons we should observe a decrease in their number due to the previously mentioned 
instanton-antiinstanton annihilation mechanism. An analogous annihilation takes place for opposite charge   $Q=\frac{1}{2}$ instantons.
The  density decrease should then drive the observed one for the string tension. Actually, our monitoring of individual 
configurations is quite consistent with this mechanism being the main cause of peak destruction. 

In summary, we have  shown that the decrease of the string tension with cooling goes 
parallel with the decrease in the number of peaks in the action density.  This shows a clear relation
between the occurence of gauge invariant structures and Confinement. We have offered an explanation in which 
the decrease in peak number is due to the annihilation  of  instanton-antiinstanton pairs. One can  explain  the
observed dependence of  the string tension  with the instanton density, within  the model  of Ref. \cite{uslat93,uscool2}.
This is a $Q=\frac{1}{2}$ instanton liquid model with density close to $(0.7 fm)^{-4}$. We have seen that the observed
total action per peak ($4 \pi^2$) is not far from the prediction of this model. Although we lack a determination
of the number of  $Q=\frac{1}{2}$ instantons present in our configurations, we have analysed individual 
configurations and verified that indeed they have a large number of the latter objects. It is possible 
that other models also explain the data. 
In any case, the association of gauge invariant structures with Confinement need  not be necessarily in contradiction
with the dual superconductor description. Indeed, we have recently observed that the   $Q=\frac{1}{2}$ instantons 
give rise, when maximally abelian projected, to monopole (both elementary and extended) currents \cite{politony}
going exactly through their centers. A monopole-instanton correlation study is in progress.   

\section*{Acknowledgements}
This work was supported by the CICYT grants AEN93-0693 and the EC network CHRX-CT93-0132. Useful conversations with M. Polikarpov are acknowledged.

\bibliography{ref}

\begin{thebibliography}{10}

\bibitem{thooft6}
G.~'t~Hooft.
\newblock {\em Nucl. Phys.}  B190 (1981)  455.

\bibitem{thooft1}
G.~'t~Hooft.
\newblock In A.~Zichichi, editor, {\em High Energy Physics, Proceedings of the
  EPS International Conference}. Editrici Compositori,  1976.

\bibitem{mandel1}
S.~Mandelstam.
\newblock {\em Phys. Rep.}  23C (1976)  245.

\bibitem{chernodub}
 M.N.~Chernodub,~M.~I.~Polikarpov  and A.~I. Veselov.
\newblock {\em Phys. Lett.}  B342 (1995)  303.

\bibitem{cool1}
B.~Berg.
\newblock {\em Phys. Lett.}  104B (1981)  475.

\bibitem{cool2}
M.~Teper.
\newblock {\em Phys. Lett.}  162B (1985)  357.

\bibitem{polikarp}
M.~I. Polikarpov and A.~I. Veselov.
\newblock {\em Nucl. Phys.}  B297 (1988)  34.

\bibitem{negele1}
M.-C.~Chu et~al.
\newblock {\em Nucl. Phys. B (Proc. Suppl.)}  34 (1994)  170.

\bibitem{gupta}
J.~Grandy and R.~Gupta.
\newblock {\em Nucl. Phys. B (Proc. Suppl.)}  42 (1995)  246.

\bibitem{michael2}
C.~Michael and P.~S. Spencer.
\newblock {\em Nucl. Phys. B (Proc. Suppl.)}  42 (1995)  261.

\bibitem{michael3}
C.~Michael and P.~S. Spencer.
\newblock Preprint hep-lat/9503018  (1995).

\bibitem{trottier}
H.~D. Trottier and R.~M. Woloshyn.
\newblock {\em Phys. Rev.}  D50 (1994)  6939.

\bibitem{uscool}
M.~Garc\'{\i}a~P\'erez et~al.
\newblock {\em Nucl. Phys.}  B413 (1994)  535.

\bibitem{uslat94}
A.~Gonz\'alez-Arroyo and P.~Mart\'{\i}nez.
\newblock {\em Nucl. Phys. B (Proc. Suppl.)}  42 (1995)  243.

\bibitem{uscoolnew}
A.~Gonz\'alez-Arroyo,~P. Mart\'{\i}nez and A.~Montero; in~preparation.

\bibitem{uslat93}
M.~Garc\'{\i}a P\'erez,~A. Gonz\'alez-Arroyo and P.~Mart\'{\i}nez.
\newblock {\em Nucl. Phys. B (Proc. Suppl.)}  34 (1994)  228.

\bibitem{uscool2}
A.~Gonz\'alez-Arroyo and P.~Mart\'{\i}nez.
\newblock {\em Investigating Yangs-Mills theory and Confinement ...}  Preprint
  FTUAM/95-15, hep-lat/9507001.

\bibitem{digiacomo}
M.Campostrini et~al.
\newblock {\em Phys. Lett.}  255B (1989)  403.

\bibitem{teperarg}
M.~Teper.
\newblock {\em Nucl. Phys.}  B411 (1994)  855.

\bibitem{us1}
M.~Garc\'{\i}a P\'erez,~A. Gonz\'{a}lez-Arroyo and B.~Soderberg.
\newblock {\em Phys. Lett.}  235B (1990)  117.

\bibitem{jouphys}
M.~Garc\'{\i}a P\'erez and A.~Gonz\'alez-Arroyo.
\newblock {\em J. Phys.}  A26 (1993)  2667.

\bibitem{politony}
A.~Gonz\'alez-Arroyo and M.~Polikarpov; in~preparation.

\end{thebibliography}

\newpage

\begin{center}

\large\bf {Figure Captions}

\end{center}

\medskip

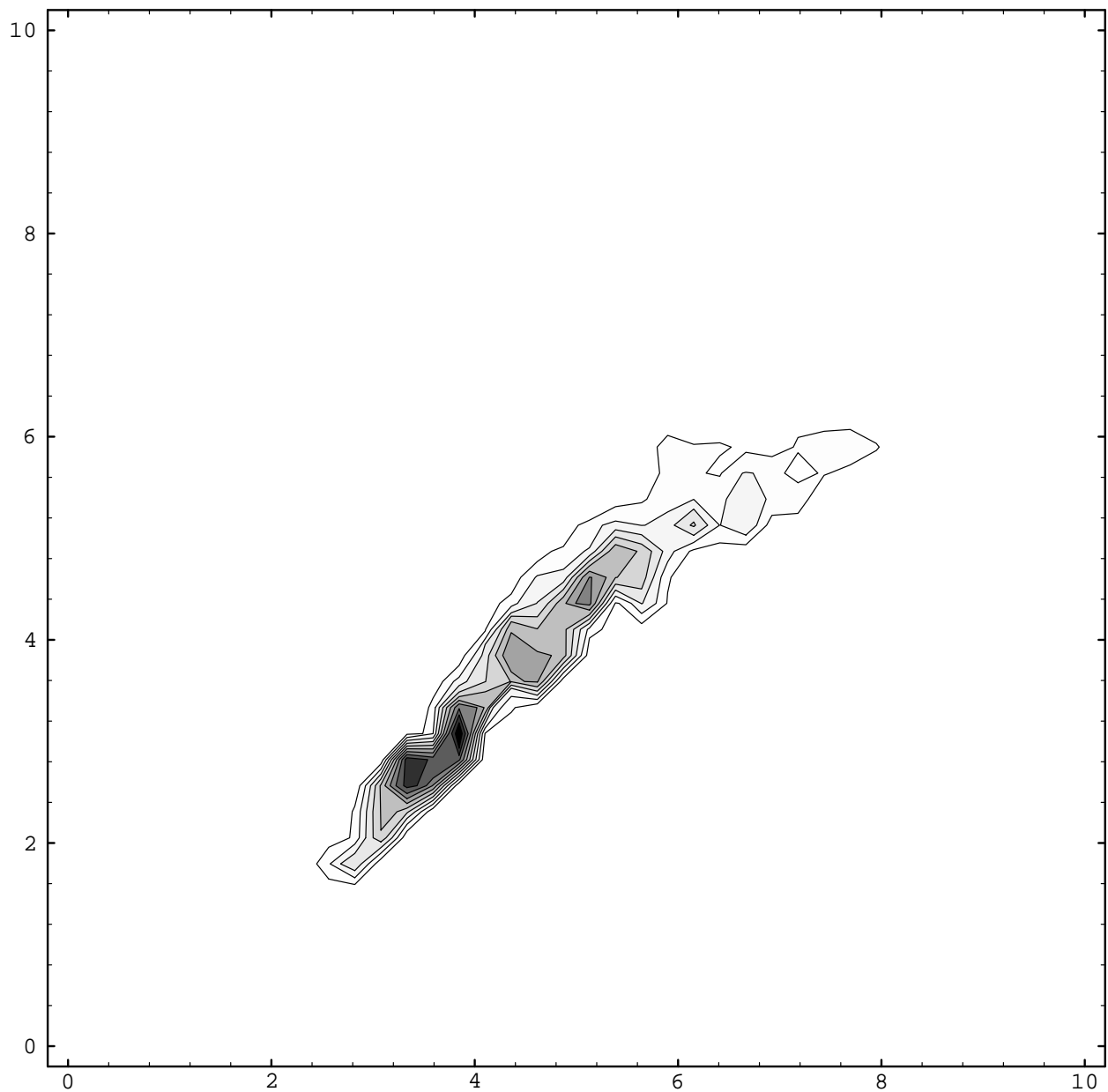
Figure 1: Contour plot which shows the combined distribution of variables $\rho_1$ (vertical axis)
 and $\rho_2$ (horizontal axis) defined in Eq. (3) for all our peaks obtained at  50
cooling steps. 

\medskip

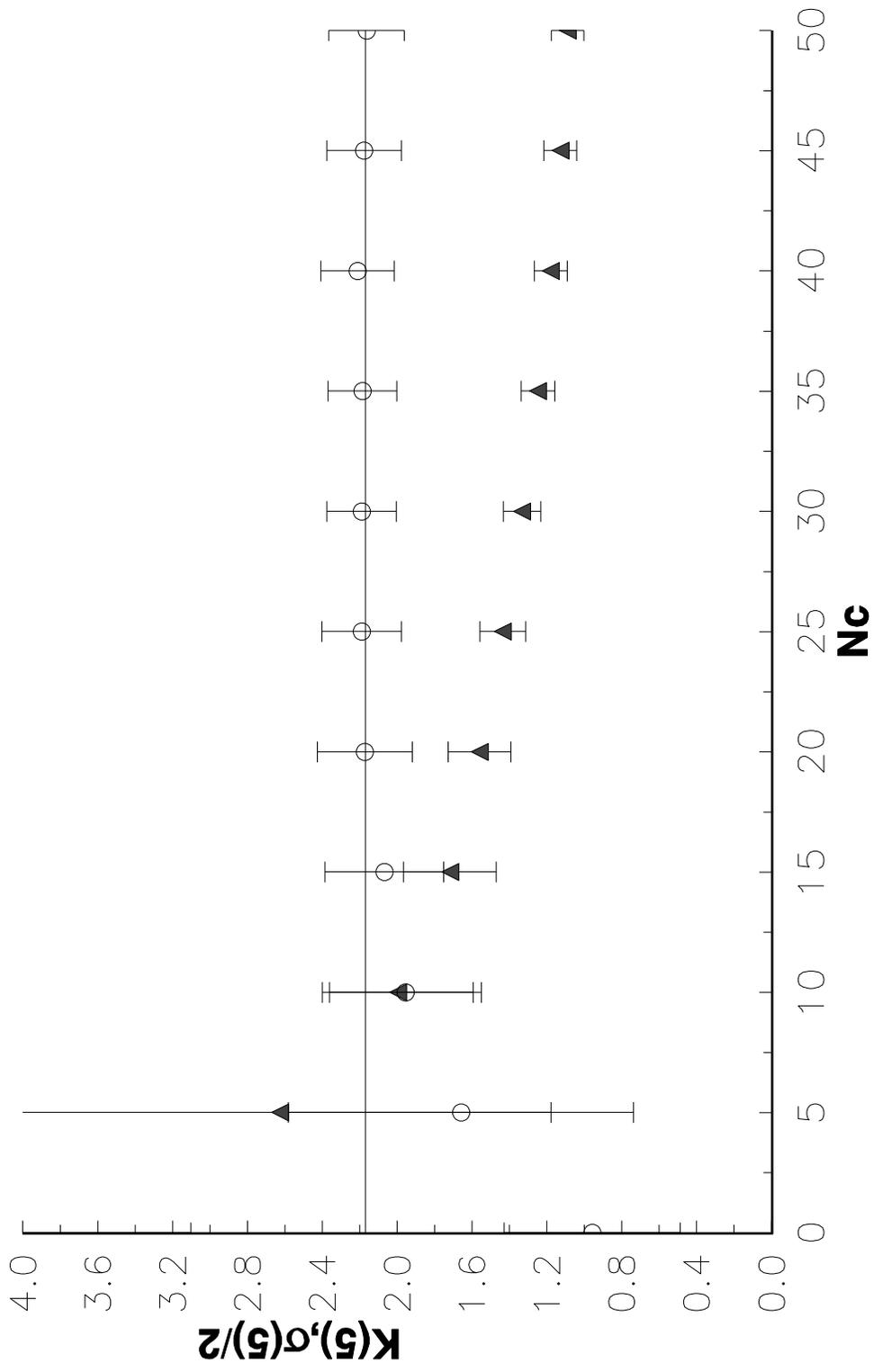
Figure 2:  The value of K (circles) defined in Eq. (4) is plotted  as a function of the number
of cooling steps  $N_c$.  The horizontal  line $ K = 2.17 $ comes from a fit to the data from 15 to 50 cooling steps.
 For comparison, in the  same plot the  value of  $\frac{\sigma}{2}$ (triangles) in $fm^{-2}$ is displayed.
In both cases the value of the string tension $\sigma$ is extracted from correlations of Polyakov loops at distances 5
and 4.
\medskip

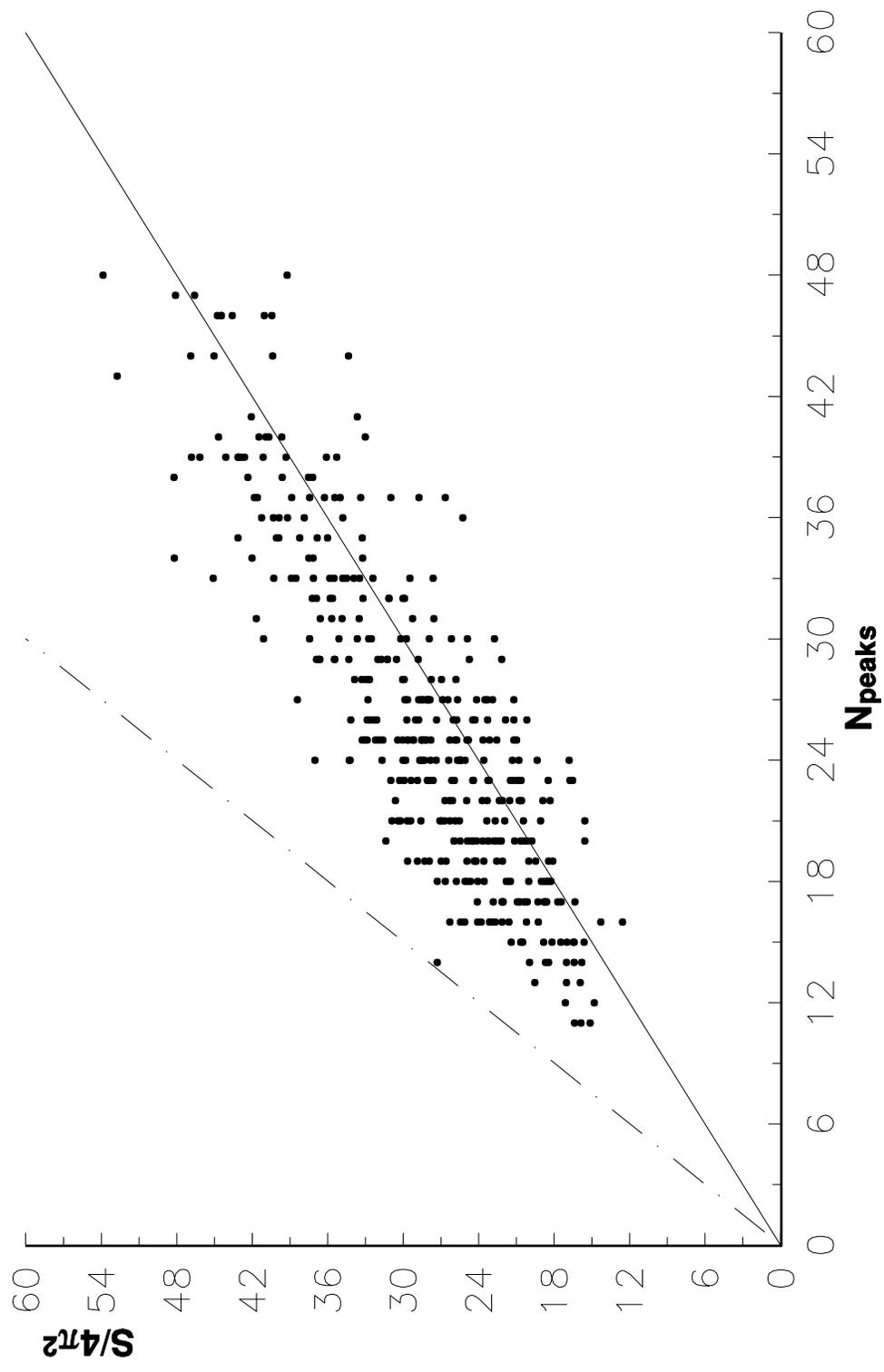
Figure 3: A density plot showing for all our configurations and all cooling steps from 20 to 50 (in steps of 5)
the total action of the configuration in units $4 \pi^2$ ($\frac{S}{ 4 \pi^2}$) versus the number of peaks $N_{peaks}$ 
of this configuration.
The two lines correspond to the predictions for a gas of $Q=1$ instantons  (slope 2)
and of $Q=\frac{1}{2}$ instantons (slope 1). 

\medskip

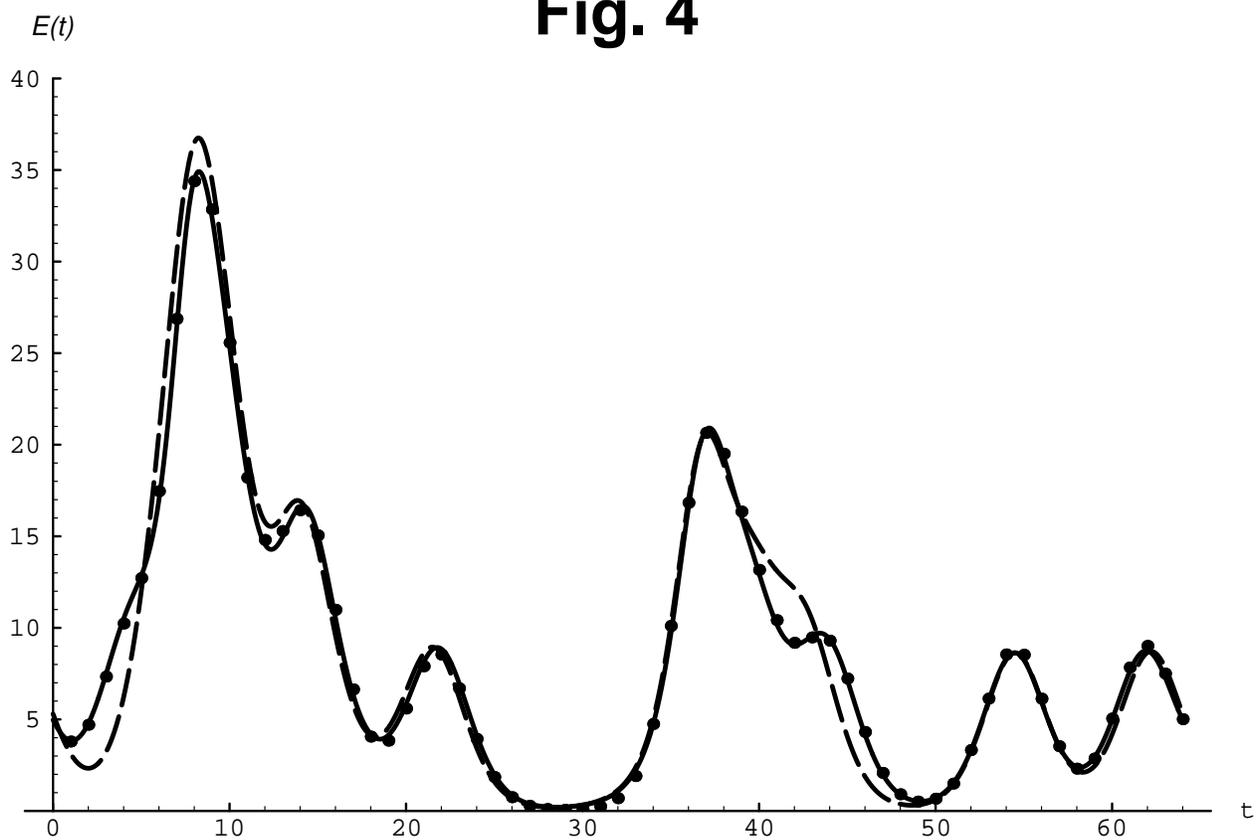
Figure 4: We show the energy profile (Eq. (6)) for one of our  configurations at 240 cooling steps. The discontinuous
line  is obtained by adding the energy profiles of a set of $Q=\frac{1}{2}$ and $Q=1$ instantons
centered at the time positions where we have observed peaks in the action density ( no free parameters). Three  out of
 the eleven peaks are  $Q=1$
instantons with values of $\rho$ which agree with the $\rho_1$ and $\rho_2$ parameters of the peak. By allowing the centers
of the instantons to vary slightly one gets the continuous line.

\newpage 

\end{document}